\documentclass[final,1p,times]{elsarticle} 
\usepackage{graphicx} 
\usepackage{amssymb} 
\usepackage{amsthm} 
\usepackage{lineno} 

\journal{Nuclear Physics A} 

\usepackage{epsfig,epsf}
\usepackage{verbatim}
\usepackage{epstopdf}
\usepackage{bm} 
\newcommand{\beq}{\begin{equation}}
\newcommand{\beql}[1]{\begin{equation}\label{#1}}
\newcommand{\eeq}{\end{equation}}
\newcommand{\bea}{\begin{eqnarray}}
\newcommand{\eea}{\end{eqnarray}}
\newcommand{\be}{\begin{eqnarray}}
\newcommand{\ee}{\end{eqnarray}}
\def\eq#1{{(\ref{#1})}}
\def\fig#1{{Fig.~\ref{#1}}}

\newcommand{\as}{\alpha_s}

\newcommand{\Lb}{\left(}
\newcommand{\Rb}{\right)}

\newcommand{\un}{\underline}

\begin{document} 

\begin{frontmatter} 


\title{Gluon saturation effects on $J/\psi$ production in heavy ion collisions }

\author{Kirill Tuchin}

\address{Department of Physics and Astronomy, Iowa State University, Ames, IA 50011\\ and \\
RIKEN BNL Research Center, Upton, NY 11973-5000}

\begin{abstract} 

We discuss a novel mechanism for $J/\psi$ production in nuclear collisions arising due to the high density of gluons. We demonstrate  that gluon saturation in the 
colliding nuclei is a dominant source of $J/\psi$ suppression and can explain its experimentally observed rapidity and centrality dependence.

\end{abstract} 

\end{frontmatter} 

\linenumbers 



\section{Introduction}

\begin{figure}[ht]
      \includegraphics[width=10cm]{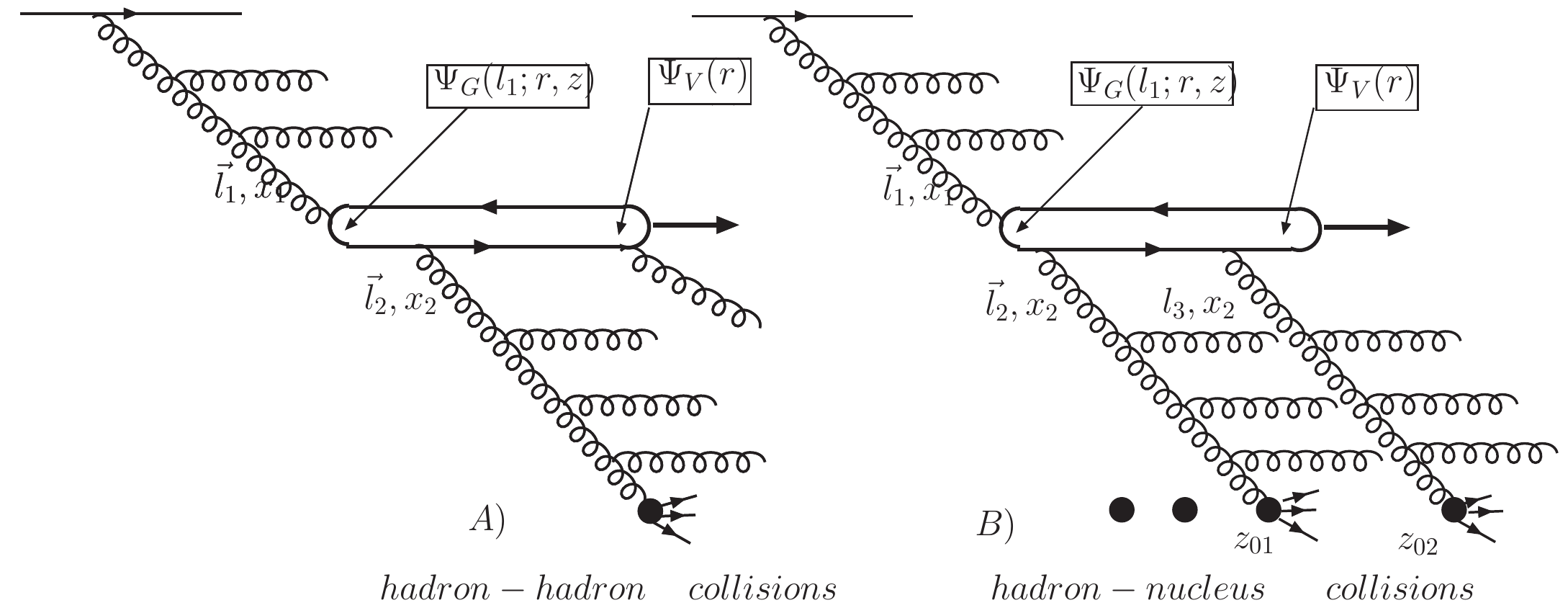} 
\caption{The  process of inclusive J/$\psi$  production in
  hadron-hadron (\protect\fig{psi1}-A) and  in hadron-nucleus
  collisions (\protect\fig{psi1}-B).  }
\label{psi1}
\end{figure}

The mechanism of $J/\psi$ production in high energy nuclear collisions is different from that in hadron-hadron collisions \cite{Kharzeev:2008cv,Kharzeev:2008nw}. Consider first the $J/\psi$ production in
hadron--hadron collisions. The leading contribution is given by the
two-gluon fusion (i) $G+G\to J/\psi + \,\mathrm{soft\,\, gluon}$, see \fig{psi1}-A. This process is of the order $\mathcal{O}(\as^5)$. The three-gluon fusion (ii) $G+G+G\to J/\psi$, see \fig{psi1}-B, is parametrically suppressed as it is proportional to
$\mathcal{O}(\as^6)$. However, in hadron-nucleus collisions an
additional gluon can be attached to the nucleus. This brings in an
additional enhancement by a factor $\sim A^{1/3}$. If the collision energy is high enough, the coherence length becomes much larger than the size of the interaction region. In this case all $A$ nucleons interact coherently as a quasi-classical field \cite{MV,JIMWLK,Kovchegov:1999yj,Kovchegov:1996ty}. In the quasi-classical 
approximation $\as^2A^{1/3}\sim 1$. Therefore, the three-gluon fusion is actually \emph{enhanced} by $1/\as$ as compared to the two-gluon fusion process. Similar conclusion holds for heavy-ion collisions (we do not consider any final state processes  leading to a possible formation of the Quark-Gluon Plasma). This approach has been previously applied in Ref.~\cite{KT} to $J/\psi$ production in $dAu$ collisions, although the nuclear geometry was oversimplified. We note a reasonable agreement of this earlier approach with the $dA$ data. 

\section{New mechanism of $J/\psi$ production}
  
\fig{psi1}-B represents the contribution of the order $(\as^2 A^{1/3})^2$. 
In general,   there must be an odd number of gluons connected to the charm fermion line because the quantum numbers  of $J/\psi$ and of gluon are $1^{--}$. Therefore, each inelastic interaction of  the $c \bar{c}$ pair must involve two
nucleons and hence is of the order $(\as^2 A^{1/3})^{2n}$, where $n=1,2,\ldots, A/2 $ is the number of nucleon pairs. To take this into account it is convenient to write the cross section as  the sum over all inelastic processes (labeled by the index
$n$). This sum  involves only even number of interactions. For a heavy nucleus $A\gg 1$ and for $N_c\gg 1$ the calculation can be significantly simplified. The final result can be written in terms of the saturation scale $Q_{s,A}^2$ \cite{GLR,MUQI,dipole}. Its value was extracted from the fit   
 of the multiplicities  of nuclear reactions at RHIC \cite{KN1,KLN} and 
 from fits of the  $F_2$ structure function in DIS
 \cite{MOD,MOD1,MOD2,MOD3}.   
 The main result reads \cite{Kharzeev:2008cv,Kharzeev:2008nw}
  \be
 \frac{d N^{A A}(Y,b)}{ d Y}&=& C \frac{d N^{pp}(Y)}{d Y}\,\,\int d^2
 s\, \,T_{A_1}(\un s)\,T_{A_2} \Lb\un{b} - \un{s}\Rb\, 
\Lb  Q^2_{s,A_1}\Lb x_1,\un{s}\Rb \,\,+\,\,Q^2_{s,A_2}\Lb x_2,\un{b} -
\un{s}\Rb\Rb \,\frac{1}{m^2_c} 
 \nonumber\\
 & &
 \times  \int^{\infty}_0\!\! d \zeta\,\,\zeta^9 \,\,K_2(\zeta)\,
 \,\,\exp\left[ - \frac{\zeta^2}{8 m^2_c}\,\Lb
 Q^2_{s,A_1}(x_1,\un{s})\,+\,Q^2_{s,A_2}(x_2, \un{b} - \un{s})\Rb 
 \right]\,,\label{IP5}
  \ee
    where $x_{1/2}=\frac{m_{\psi}}{\sqrt{s}}\,e^{\mp Y}$.
Eq.~\eq{IP5} is derived in the quasi-classical approximation which takes into account multiple scattering of the $c\bar c$ pair in the cold nuclear medium. At forward rapidities at RHIC and at LHC the gluon distribution function evolves according to the evolution equations of the color glass condensate. Inclusion of this evolution in the case of $J/\psi$ production presents a formidable technical challenge. Therefore we adopt a  phenomenological approach of \cite{KLN,Kharzeev:2004yx}  in which the quantum evolution is encoded in the energy/rapidity dependence of the saturation scale.

In our numerical calculations we take explicit account of the impact parameter dependence of the saturation scales of each nucleus. We employ the Glauber  approximation and assume that the nucleon radius is much smaller than the   
nucleus radius.  The overall normalization constant $C$  includes the color  and the geometric factors $C^2_F/(4 \pi^2 \as S_p)$ where $S_p$ 
is interaction area in proton--proton collisions. $C$ also  includes
the amplitude of quark--antiquark annihilation into $J/\psi$ and a
soft gluon in the case of $pp$ collisions. This
amplitude as well as the  mechanism of \fig{psi1}-A  have a significant theoretical uncertainty.  
Therefore, we decided to parameterize these contributions by an
overall normalization constant in  \eq{IP5}.

\begin{figure}[ht]
\begin{tabular}{cc}
      \includegraphics[width=6.5cm]{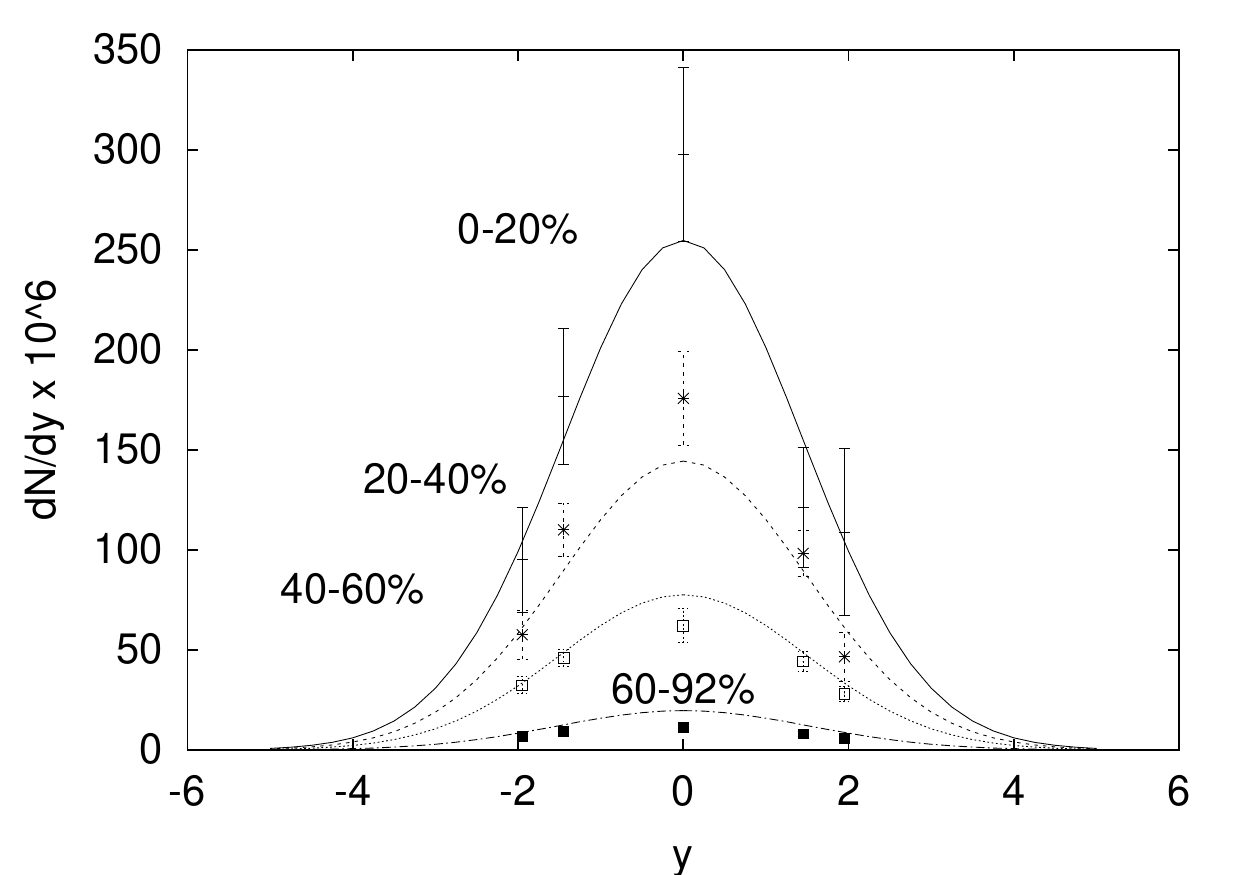} &
      \includegraphics[width=6.5cm]{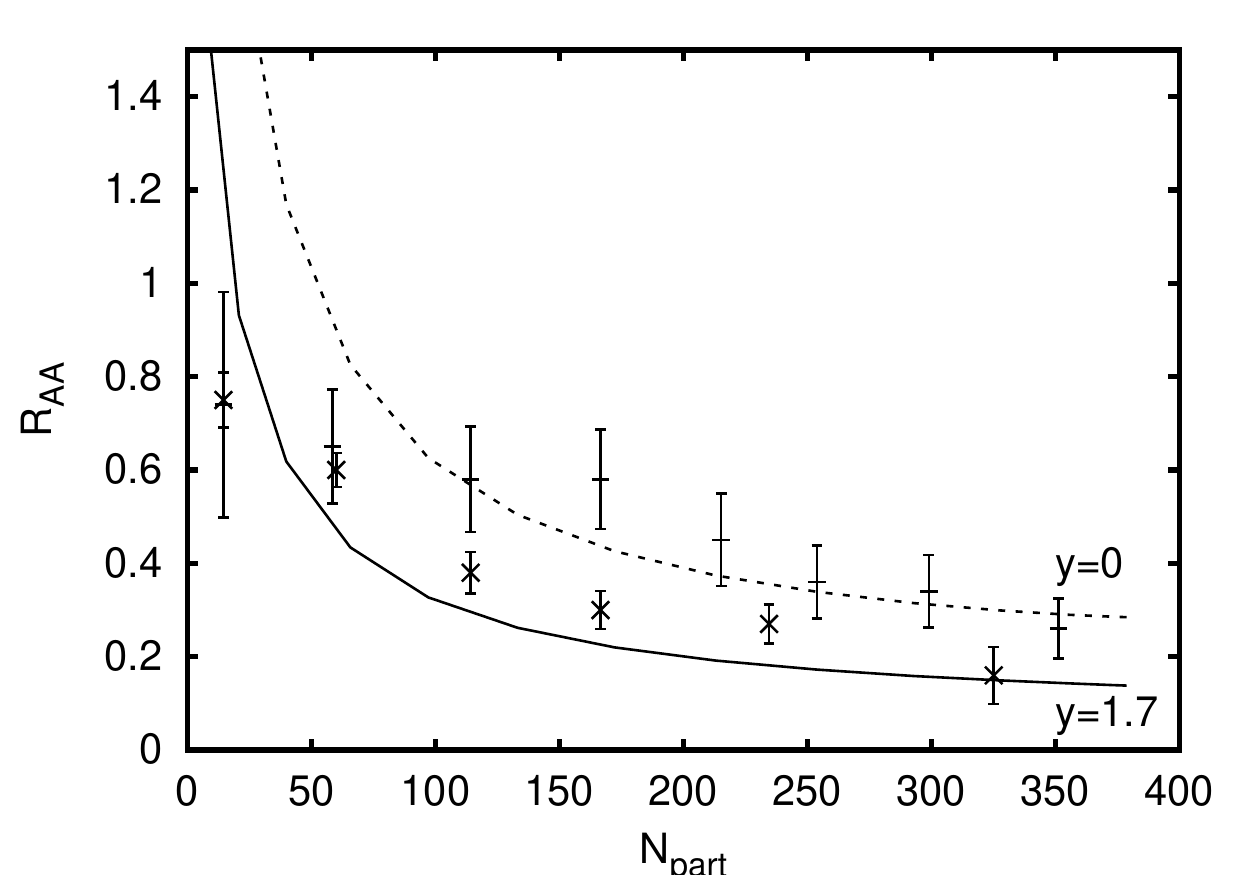}\\
      (a) & (b)
 \end{tabular}     
\caption{(a) $J/\psi$ rapidity distribution in Au-Au collisions for
  different centrality cuts. (b) Nuclear modification factor for $J/\psi$ production in heavy-ion collisions for different rapidities. Experimental data from \cite{Adare:2006ns}.}
\label{fig:BdNdy}
\end{figure}

The  rapidity distribution of $J/\psi$'s in $pp$ collisions, the
factor ${d N^{pp}}/{d Y}$ appearing in Eq. \ref{IP5}, is fitted to
the experimental data given in \cite{Adare:2006kf} with a single
gaussian. In figure \ref{fig:BdNdy}(a)
the results provided by Eq.~\eq{IP5} are then compared
to the experimental data from PHENIX Collaboration \cite{Adare:2006ns} for
Au-Au collisions at $\sqrt{s}=200$ GeV. The global normalization factor $C$ is
found from the overall fit. There are no other free
parameters.  The agreement of the theoretical results
with experimental data is reasonable. It is evident that the effect of the gluon saturation on the  $J/\psi$ rapidity distribution in nucleus-nucleus collisions is to make its width a decreasing function of centrality. The distribution in the most central bin is significantly  more narrow than in the  peripheral bin.

It is important that we describe well the data in the semi-peripheral region. This ensures that our model gives a good description of the $J/\psi$ production in $dAu$ collisions. We also note that an earlier approach \cite{KT} in which the same model was employed (albeit with an oversimplified nuclear geometry) provided a reasonable description of the data.  Still a more detailed investigation is required which takes into account the exact deuteron and gold nuclear distributions. This will allow a model-independent fixing of the overall normalization constant $C$. We plan to present such an analysis in the near future. 

To emphasize the nuclear dependence of the inclusive cross sections it is convenient to introduce the nuclear modification factor defined in a usual way. 
 In \fig{fig:BdNdy}(b) we plot the result of our calculation. The nuclear modification factor exhibits the following  two important features: (i) unlike the open charm production, $J/\psi$ is suppressed even at $y=0$;  (ii) cold nuclear matter effects account for a significant part of the ``anomalous" $J/\psi$ suppression in heavy-ion collisions both at $y=0$ and $y=2$.


\section{Conclusions}\label{concl}

The main results of this paper are exhibited in Fig.~\ref{fig:BdNdy}. 
It is seen that the rapidity and centrality dependence of $J/\psi$ production are reproduced with a reasonable accuracy even without taking into account any hot nuclear medium effects. 
This observation allows to conclude that a fair amount (and perhaps most) of the $J/\psi$ suppression in high energy heavy-ion collisions arises from the cold nuclear matter effects. In other words, $J/\psi$ is expected to be strongly suppressed even if there were no hot nuclear matter produced. 

The reason for $J/\psi$ suppression at mid-rapidities is that the multiple scattering of $c\bar c$ in the cold nuclear medium increases the relative momentum between the quark and antiquark, which makes the bound state formation less probable. It was proven in \cite{KKT} that unless quantum $\log(1/x)$ corrections become important, the inclusive gluon production satisfies the sum rule that requires the nuclear modification factor to be of order unity. Similar sum rule holds for heavy quark production but fails in the case of a 
bound states, such as $J/\psi$.

We realize that although our calculation gives the parametrically leading result at high gluon density, other production channels involving the gluon radiation in the final state and the color octet mechanism of  $J/\psi$ production may give phenomenologically significant contributions. These are likely to become the leading mechanisms in the peripheral collisions where the strength of the gluon fields is significantly diminished. However, we believe that our main results are robust for central high-energy collisions of heavy ions.


\section*{Acknowledgments} 
I am grateful to D.~Kharzeev, E.~Levin and  M.~Nardi for a fruitful collaboration. 
This work was supported in part by the U.S. Department of Energy under Grant No. DE-FG02-87ER40371; I would like to
thank RIKEN, BNL,
and the U.S. Department of Energy (Contract No. DE-AC02-98CH10886) for providing facilities essential
for the completion of this work.  RBRC-797.

\end{document}